\documentstyle[12pt,amsfonts]{article}
\newcommand{\swav}[1]{ \stackrel{\sim}{#1} }

\newcommand{\ehat}{ \hat U_{\epsilon} }

\newcommand{\define}{ \stackrel{\triangle}{=} }

\def\be{\begin{equation}}
\def\ee{\end{equation}}
\def\ba{\begin{array}}
\def\ea{\end{array}}

\begin{document}
\title{\bf Path Integral Quantization of Quantum Gauge General Relativity }
\author{{Ning Wu}
\thanks{email address: wuning@mail.ihep.ac.cn}
\\
\\
{\small Institute of High Energy Physics, P.O.Box 918-1,
Beijing 100039, P.R.China}}
\maketitle
\vskip 0.8in

~~\\
PACS Numbers: 11.15.-q, 04.60.-m, 04.20.Cv, 11.10.Gh. \\
Keywords: general relativity, gauge field,
        quantum gravity, path integral quantization, Feynman rules .\\

\vskip 0.4in

\begin{abstract}

Path integral quantization of quantum gauge general relativity is
discussed in this paper. First, we deduce the generating functional
of green function with external fields. Based on this generating
functional, the propagators of gravitational gauge field and related
ghost field are deduced. Then, we calculate Feynman rules of various
interaction vertices of three or four gravitational gauge fields and vertex
between ghost field and gravitational gauge field. Results in this
paper are the bases of calculating vacuum polarization of gravitational
gauge field and vertex correction of gravitational couplings in
one loop diagram level. As we have pointed out in previous paper,
quantum gauge general relativity is perturbative renormalizable, and
a formal proof on its renormalizability is also given in the previous
paper. Next step, we will calculate one-loop and two-loop renormalization
constant, and to prove that the theory is renormalizable in one-loop
and two-loop level by direct calculations.
\\
\end{abstract}

\Roman{section}

\section{Introduction}
\setcounter{equation}{0}

Quantum gravity is proposed to unify general relativity and quantum theory.
One of the biggest troubles for quantum gravity is the problems of
perturbative renormalization. Gauge gravity is studied for a long time,
and there are many versions of gauge gravity\cite{a01,a02,06,a04}. It is
expected that gauge gravity could solve the problem of renormalization of
quantum gravity. \\

Quantum gauge general relativity is
proposed to solve this problem\cite{01,02,03,04,05,06,07}.
It is a quantum theory of gravity
proposed in the framework of quantum gauge field theory.
In 2003,  Quantum Gauge General Relativity(QGGR)
is proposed in the framework of QGTG.
Unlike Einstein's general theory of relativity, the cornerstone
of QGGR is the gauge principle, not the principle of equivalence,
which will cause far-reaching influence to the theory of gravity.
In QGGR, the field equation of gravitational gauge field is just
the Einstein's field equation, so in classical level, we can set up
its geometrical formulation\cite{15}, and QGGR returns to
Einstein's general relativity in classical level.  The
field equation of gravitational gauge field in QGGR is the same as
Einstein's field equation in general relativity, so two equations
have the same solutions, though mathematical expressions of the
two equations are completely different.
For classical tests of gravity, QGGR gives out the same theoretical
predictions as those of GR\cite{16}, and for non-relativistic
problems, QGGR can return to Newton's classical theory
of gravity\cite{17}. Based  on the coupling
between the spin of a particle and gravitoelectromagnetic
field, the equation of motion of spin can be obtained in QGGR.
In post Newtonian approximations, this equation of motion of
spin gives out the same results as those of GR\cite{18}.
The equation of motion of a spinning test particle in gravitational
field can also obtained\cite{18a}. It's found that this motion
deviates from traditional geodesic curve, and the deviation effects
is detectable\cite{18b}, which can be regarded as a new classical
tests of gravity theory. QGGR is a perturbatively
renormalizable quantum theory, and based on it, quantum effects of
gravity\cite{19,20,21,22} and gravitational interactions of some basic
quantum fields \cite{23,24} can be explored. Unification of fundamental
interactions including gravity can be fulfilled in a semi-direct
product gauge group\cite{25,26,27,28}. If we use the mass
generation mechanism which is proposed
in literature \cite{29,30}, we can propose a new
theory on gravity which contains massive graviton and
the introduction of massive graviton does not affect
the strict local gravitational gauge symmetry of the
action and does not affect the traditional long-range
gravitational force\cite{31}. The existence of massive graviton
will help us to understand the possible origin of dark matter.
\\

In literature
\cite{06}, a formal proof on the renormalizability of quantum gauge general
relativity is given. The proof is not based on the calculation of loop
diagrams, but based on generalized  BRST symmetry and generalized
Ward-Takahashi identities. This case is similar to that of traditional
gauge field theory. We know that traditional gauge field theory
is a renormalizable quantum theory\cite{c01,c02,c03,c04,c05,c06}.
In gauge field theory, though there are many divergences in loop diagram
calculations, the constraints from gauge symmetry will make all
divergences cancel each other.  \\

Now, we want ask that the divergence cancellation mechanism in quantum gauge
general relativity is really work in one- or two-loop level, as what we
expected in the literature \cite{06}? In order to prove
that quantum gauge general relativity is perturbatively renormalizable
in one-loop and two-loop level, we need first to calculate propagators of
gravitational gauge field and ghost field, to determine the Feynman rules
of various interaction vertices, and to calculate all divergent one-loop
and two-loop Feynman diagrams. As a first step, we discuss quantization
of quantum gauge general relativity, and determine Feynman rules of various
vertices, which is the main goal of this paper. Next step, we will calculate
all divergent one-loop Feynman diagram and discuss the renormalization problem
of quantum gauge general relativity in one-loop level. Finally, we discuss
the renormalization problem in two-loop level. So this paper is the first
one of a serial of papers on the renormalization of quantum gauge general
relativity. All these calculations are
extremely complicated and time consuming. In order to avoid possible
mistakes in analytical deductions, all important results are calculated at
least two times, and two calculations are completely independent. Some
important results are also checked by using Mathematica. How to use
Mathematica to perform these calculations will be discussed in another
paper.
\\

\section{Quantum Gauge General Relativity}
\setcounter{equation}{0}

In quantum gauge general relativity, the most fundamental
quantity is gravitational gauge field $C_{\mu}(x)$,which
is a vector in the corresponding Lie algebra.
$C_{\mu}(x)$ can be expanded as
\be \label{2.1}
C_{\mu}(x) = C_{\mu}^{\alpha} (x) \hat{P}_{\alpha},
~~~~~~(\mu, \alpha = 0,1,2,3)
\ee
where $C_{\mu}^{\alpha}(x)$ is the component field and
$\hat{P}_{\alpha} = -i \frac{\partial}{\partial x^{\alpha}}$
is the  generator of global gravitational gauge group.
The gravitational gauge covariant derivative is given by
\be \label{2.2}
D_{\mu} = \partial_{\mu} - i g C_{\mu} (x)
= G_{\mu}^{\alpha} \partial_{\alpha},
\ee
where $g$ is the gravitational coupling constant and matrix $G$ is given by
\be \label{2.3}
G = (G_{\mu}^{\alpha}) = ( \delta_{\mu}^{\alpha} - g C_{\mu}^{\alpha} ).
\ee
Its inverse matrix is
\be \label{2.4}
G^{-1} = \frac{1}{I - gC} = (G^{-1 \mu}_{\alpha}).
\ee
Using matrix $G$ and $G^{-1}$, we can define two important
composite operators
\be \label{2.5}
g^{\alpha \beta} = \eta^{\mu \nu}
G^{\alpha}_{\mu} G^{\beta}_{\nu},
\ee
\be \label{2.6}
g_{\alpha \beta} = \eta_{\mu \nu}
G_{\alpha}^{-1 \mu} G_{\beta}^{-1 \nu}.
\ee
In quantum gauge general relativity,
space-time is always flat and space-time metric
is always Minkowski metric, so $g^{\alpha\beta}$ and $g_{\alpha\beta}$
are no longer space-time metric. They are only two composite operators
which  consist of gravitational gauge field. \\

The  field strength of gravitational gauge field is defined by
\be \label{2.7}
F_{\mu\nu} (x) \define \frac{1}{-ig} \lbrack D_{\mu}~~,~~D_{\nu} \rbrack =
F_{\mu\nu}^{\alpha}(x) \cdot \hat{P}_{\alpha}
\ee
where
\be \label{2.8}
F_{\mu\nu}^{\alpha} =
G_{\mu}^{\beta} \partial_{\beta} C_{\nu}^{\alpha}
-G_{\nu}^{\beta} \partial_{\beta} C_{\mu}^{\alpha}.
\ee
\\

The Lagrangian
of the  quantum gauge general relativity is selected to be
\be \label{2.9}
{\cal L} = ( {\rm det} G^{-1} ) {\cal L}_0,
\ee
where
\be \label{2.10}
{\cal L}_0  =  - \frac{1}{16} \eta^{\mu \rho}
\eta^{\nu \sigma} g_{\alpha \beta}
F^{\alpha}_{\mu \nu} F^{\beta}_{\rho \sigma}
 - \frac{1}{8} \eta^{\mu \rho}
G^{-1 \nu}_{\beta} G^{-1 \sigma}_{\alpha}
F^{\alpha}_{\mu \nu} F^{\beta}_{\rho \sigma}
+ \frac{1}{4} \eta^{\mu \rho}
G^{-1 \nu}_{\alpha} G^{-1 \sigma}_{\beta}
F^{\alpha}_{\mu \nu} F^{\beta}_{\rho \sigma}.
\ee
Its space-time integration gives out the action of the system
\be \label{2.101}
S = \int {\rm d}^4 x  {\cal L}.
\ee
\\

\section{Path Integral Quantization of Gravitational\\ Gauge Fields}
\setcounter{equation}{0}

Gravitational gauge field $C_{\mu}^{\alpha}$ has $4 \times 4 = 16$
degrees of freedom. But, if gravitons are massless, the system has only
$2 \times 4 = 8$ degrees of freedom. There are gauge degrees
of freedom in the theory. Because only physical degrees of
freedom can be quantized, in order to quantize the system, we
have to introduce gauge conditions to eliminate un-physical degrees
of freedom. For the sake of convenience, we take temporal gauge
conditions
\be
C_0^{\alpha} = 0, ~~~(\alpha = 0,1,2,3).
\label{10.3}
\ee
\\

In temporal gauge, the generating functional $W\lbrack J \rbrack$ is
given by
\be
W\lbrack J \rbrack = N \int \lbrack {\cal D} C\rbrack
\left(\prod_{\alpha, x} \delta( C_0^{\alpha}(x))\right)
exp \left\lbrace i \int {\rm d}^4 x ( {\cal L}
+ J^{\mu}_{\alpha} C^{\alpha}_{\mu}),
\right\rbrace
\label{10.4}
\ee
where $N$ is the normalization constant, $J^{\mu}_{\alpha}$
is a fixed external source and $ \lbrack {\cal D} C\rbrack $ is the
integration measure,
\be
\lbrack {\cal D} C\rbrack
= \prod_{\mu=0}^{3} \prod_{\alpha= 0}^{3} \prod_j
\left(\varepsilon {\rm d}C_{\mu}^{\alpha} (\tau_j)
/ \sqrt{2 \pi i \hbar} \right).
\label{10.5}
\ee
We use this generation functional as our starting
point of the path integral quantization of gravitational gauge field. \\

Generally speaking, the action of the system has local gravitational gauge
symmetry, but the gauge condition has no local gravitational gauge
symmetry. If we make a local gravitational gauge transformations,
the action of the system is kept unchanged while gauge condition will
be changed. Therefore, through local gravitational gauge transformation,
we can change one gauge condition into another gauge condition. The most
general gauge condition is
\be
f^{\alpha} (C(x)) - \varphi^{\alpha} (x) = 0,
\label{10.6}
\ee
where $\varphi^{\alpha}(x)$ is an arbitrary space-time function.
The Fadeev-Popov determinant $\Delta_f(C)$  is defined by
\be
\Delta_f^{-1} (C) \equiv
\int \lbrack {\cal D} g \rbrack
\prod_{x, \alpha} \delta \left(f^{\alpha} ( ^gC(x))
- \varphi^{\alpha}(x) \right),
\label{10.7}
\ee
where $g$ is an element of gravitational gauge group, $^gC$ is the gravitational
gauge field after gauge transformation $g$ and
$\lbrack {\cal D} g \rbrack $ is the integration measure on
gravitational gauge group
\be
\lbrack {\cal D} g \rbrack
= \prod_x {\rm d}^4 \epsilon(x),
\label{10.8}
\ee
where $\epsilon (x)$ is the transformation parameter of $\ehat$.
Both $\lbrack {\cal D} g \rbrack $ and $\lbrack {\cal D} C \rbrack $
are not invariant under gravitational gauge transformation. Suppose
that,
\be
\lbrack {\cal D} (gg') \rbrack
= J_1(g') \lbrack {\cal D} g \rbrack,
\label{10.9}
\ee
\be
\lbrack {\cal D} ~^gC \rbrack
= J_2(g) \lbrack {\cal D} C \rbrack.
\label{10.10}
\ee
$J_1(g)$ and $J_2(g)$ satisfy the following relations
\be
J_1(g) \cdot J_1(g^{-1}) = 1,
\label{10.11}
\ee
\be
J_2(g) \cdot J_2(g^{-1}) = 1.
\label{10.12}
\ee
It can be proved that, under gravitational gauge transformations,
the Fadeev-Popov determinant transforms as
\be
\Delta_f^{-1} ( ^{g'}C ) = J_1^{-1}(g') \Delta_f^{-1} (C).
\label{10.13}
\ee
\\

Insert eq.(\ref{10.7}) into eq.(\ref{10.4}), we get
\be
\begin{array}{rcl}
W\lbrack J \rbrack &=& N \int \lbrack {\cal D} g \rbrack
\int \lbrack {\cal D} C\rbrack~~
\left\lbrack \prod_{\alpha, y}
\delta( C_0^{\alpha}(y)) \right\rbrack \cdot
\Delta_f  (C) \\
&&\\
&&\cdot \left\lbrack \prod_{\beta, z}
\delta ( f^{\beta} ( ^gC(z))- \varphi^{\beta}(z) )\right\rbrack
 \cdot exp \left\lbrace i \int {\rm d}^4 x ( {\cal L}
+ J^{\mu}_{\alpha} C^{\alpha}_{\mu})
\right\rbrace .
\end{array}
\label{10.14}
\ee
Make a gravitational gauge transformation,
\be
C(x)~~ \to ~~^{g^{-1}} C(x),
\label{10.15}
\ee
then,
\be
^g C(x)~~ \to
~~^{gg^{-1}} C(x).
\label{10.16}
\ee
After this transformation, the generating functional is changed
into
\be
\begin{array}{rcl}
W\lbrack J \rbrack &=& N \int \lbrack {\cal D} g \rbrack
\int \lbrack {\cal D} C\rbrack~~
J_1(g) J_2(g^{-1}) \cdot \left\lbrack \prod_{\alpha, y}
\delta( ^{g^{-1}}C_0^{\alpha}(y)) \right\rbrack \cdot
\Delta_f  (C) \\
&&\\
&&\cdot \left\lbrack \prod_{\beta, z}
\delta ( f^{\beta} ( C(z))- \varphi^{\beta}(z) )\right\rbrack
 \cdot exp \left\lbrace i \int {\rm d}^4 x ( {\cal L}
+ J^{\mu}_{\alpha} \cdot  ^{g^{-1}} \!\!\! C^{\alpha}_{\mu})
\right\rbrace.
\end{array}
\label{10.17}
\ee
\\

Suppose that the gauge transformation $g_0(C)$ transforms general
gauge condition $f^{\beta}(C) - \varphi^{\beta} = 0$ to temporal
gauge condition $C_0^{\alpha} = 0$, and suppose that this transformation
$g_0(C)$ is unique. Then two $\delta$-functions in eq.(\ref{10.17}) require
that the integration on gravitational gauge group must be in the
neighborhood of $g^{-1}_0(C)$. Therefore eq.(\ref{10.17}) is changed into
\be
\begin{array}{rcl}
W\lbrack J \rbrack &=& N \int \lbrack {\cal D} C\rbrack~~
\Delta_f  (C)  \cdot \left\lbrack \prod_{\beta, z}
\delta ( f^{\beta} ( C(z))- \varphi^{\beta}(z) )\right\rbrack \\
&&\\
&& \cdot exp \left\lbrace i \int {\rm d}^4 x ( {\cal L}
+ J^{\mu}_{\alpha} \cdot  ^{g_0}\!C^{\alpha}_{\mu})
\right\rbrace \\
&&\\
&& \cdot J_1(g_0^{-1}) J_2(g_0) \cdot \int \lbrack {\cal D} g \rbrack
\left\lbrack \prod_{\alpha, y}
\delta( ^{g^{-1}}C_0^{\alpha}(y))\right\rbrack.
\end{array}
\label{10.18}
\ee
The last line in eq.(\ref{10.18}) will cause no trouble in renormalization,
and if we consider the contribution from ghost fields which will
be introduced below, it will become a quantity which is independent
of gravitational gauge field. So, we put it into normalization constant
$N$ and still denote the new normalization constant as $N$. We also
change $J^{\mu}_{\alpha} ~  ^{g_0}\!C^{\alpha}_{\mu}$
into $J^{\mu}_{\alpha} C^{\alpha}_{\mu}$, this will cause no
trouble in renormalization. Then we get
\be
\begin{array}{rcl}
W\lbrack J \rbrack &=& N \int \lbrack {\cal D} C\rbrack~~
\Delta_f  (C)  \cdot \lbrack \prod_{\beta, z}
\delta ( f^{\beta} ( C(z))- \varphi^{\beta}(z) )\rbrack \\
&&\\
&& \cdot exp \lbrace i \int {\rm d}^4 x ( {\cal L}
+ J^{\mu}_{\alpha} C^{\alpha}_{\mu}) \rbrace.
\end{array}
\label{10.19}
\ee
In fact, we can use this formula as our start-point of path integral
quantization of gravitational gauge field, so we need not worried
about the influences of the third line in eq.(\ref{10.18}).
\\

Use another functional
\be
exp \left\lbrace - \frac{i}{2 \alpha}
\int {\rm d}^4 x \eta_{\alpha \beta}
\varphi^{\alpha}(x) \varphi^{\beta}(x) \right\rbrace,
\label{10.20}
\ee
times both sides of eq.(\ref{10.19}) and then make functional
integration
$\int \lbrack {\cal D} \varphi  \rbrack$,
we get
\be
W\lbrack J \rbrack  = N \int \lbrack {\cal D} C\rbrack~~
\Delta_f (C) \cdot exp \left\lbrace i \int {\rm d}^4 x ( {\cal L}
- \frac{1}{2 \alpha} \eta_{\alpha \beta} f^{\alpha} f^{\beta}
+ J^{\mu}_{\alpha} C^{\alpha}_{\mu}) \right\rbrace.
\label{10.21}
\ee
Now, let's discuss the contribution from $\Delta_f (C)$ which
is related to the ghost fields. Suppose that $g = \ehat$
is an infinitesimal gravitational gauge transformation.
The gravitational gauge transformation of gravitational gauge
field $C_{\mu}^{\alpha} (x)$ is\cite{01,02,03,04,05,06,07}
\be
C_{\mu}^{\alpha}(x) \to ^gC_{\mu}^{\alpha}(x)=
\Lambda ^{\alpha}_{~~\beta} (\ehat C_{\mu}^{\beta}(x))
- \frac{1}{g} (\ehat  \partial_{\mu} \epsilon^{\alpha}(y)),
\label{10.21a}
\ee
Then we have
\be
^gC_{\mu}^{\alpha} (x)
= C_{\mu}^{\alpha} (x)
- \frac{1}{g} {\mathbf D}_{\mu~\sigma}^{\alpha} \epsilon^{\sigma},
\label{10.22}
\ee
where
\be
{\mathbf D}_{\mu~\sigma}^{\alpha}
=\delta^{\alpha}_{\sigma} \partial_{\mu}
- g \delta^{\alpha}_{\sigma} C_{\mu}^{\beta} \partial_{\beta}
+ g \partial_{\sigma} C_{\mu}^{\alpha}.
\label{10.23}
\ee
In order to deduce eq.(\ref{10.22}), the following relation is used
\be
\Lambda^{\alpha}_{~\beta}
= \delta^{\alpha}_{\beta}
+ \partial_{\beta} \epsilon^{\alpha}
+ o( \epsilon^2).
\label{10.24}
\ee
${\mathbf D}_{\mu}$ can be regarded as the covariant derivative
in adjoint representation, for
\be
{\mathbf D}_{\mu} \epsilon
= \lbrack D_{\mu} ~~~,~~~ \epsilon \rbrack,
\label{10.25}
\ee
\be
({\mathbf D}_{\mu} \epsilon)^{\alpha}
= {\mathbf D}_{\mu~\sigma}^{\alpha} \epsilon^{\sigma}.
\label{10.26}
\ee
Using all these relations,  we have,
\be
f^{\alpha} (^gC(x)) = f^{\alpha} (C)
- \frac{1}{g} \int {\rm d}^4 y
\frac{\delta f^{\alpha}(C(x))}{\delta C_{\mu}^{\beta}(y)}
{\mathbf D}_{\mu~\sigma}^{\beta}(y) \epsilon^{\sigma}(y)
+ o(\epsilon^2).
\label{10.27}
\ee
Therefore, according to eq.(\ref{10.7}) and eq.(\ref{10.6}), we get
\be
\Delta_f^{-1} (C) =
\int \lbrack {\cal D} \epsilon \rbrack
\prod_{x, \alpha}
\delta \left( - \frac{1}{g}     \int {\rm d}^4 y
\frac{\delta f^{\alpha}(C(x))}{\delta C_{\mu}^{\beta}(y)}
{\mathbf D}_{\mu~\sigma}^{\beta}(y) \epsilon^{\sigma}(y) \right).
\label{10.28}
\ee
Define
\be
\begin{array}{rcl}
{\mathbf M}^{\alpha}_{~\sigma}(x,y) &=& -g
\frac{\delta}{\delta \epsilon^{\sigma}(y)}
f^{\alpha}(^gC(x)) \\
&&\\
&=&\int {\rm d}^4 z
\frac{\delta f^{\alpha}(C(x))}{\delta C_{\mu}^{\beta}(z)}
{\mathbf D}_{\mu~\sigma}^{\beta}(z) \delta(z-y) .
\end{array}
\label{10.29}
\ee
Then eq.(\ref{10.28}) is changed into
\be
\begin{array}{rcl}
\Delta_f^{-1} (C) &=&
\int \lbrack {\cal D} \epsilon \rbrack
\prod_{x, \alpha}
\delta \left( - \frac{1}{g} \int {\rm d}^4 y
{\mathbf M}^{\alpha}_{~\sigma}(x,y) \epsilon^{\sigma}(y)
\right)  \\
&&\\
&=& const. \times (det {\mathbf M} )^{-1}.
\end{array}
\label{10.30}
\ee
Therefore,
\be
\Delta_f (C) = const. \times det {\mathbf M}.
\label{10.31}
\ee
Put the above constant into normalization constant, then generating
functional eq.(\ref{10.21}) is changed into
\be
W\lbrack J \rbrack  = N \int \lbrack {\cal D} C\rbrack~~
det {\mathbf M} \cdot exp \left\lbrace i \int {\rm d}^4 x ( {\cal L}
- \frac{1}{2 \alpha} \eta_{\alpha \beta} f^{\alpha} f^{\beta}
+ J^{\mu}_{\alpha} C^{\alpha}_{\mu}) \right\rbrace.
\label{10.32}
\ee
\\

In order to evaluate the contribution from $det {\mathbf M}$,
we introduce ghost fields $\eta^{\alpha}(x)$
and $\bar{\eta}_{\alpha}(x)$. Using the following relation
\be
\int \lbrack {\cal D} \eta \rbrack
\lbrack {\cal D}\bar{\eta} \rbrack
exp \left\lbrace i \int {\rm d}^4 x {\rm d}^4 y ~
\bar{\eta}_{\alpha} (x) {\mathbf M}^{\alpha}_{~\beta}(x,y)
\eta^{\beta} (y) \right\rbrace
 = const. \times det {\mathbf M}
\label{10.33}
\ee
and put the constant into the normalization constant, we can get
\be
W\lbrack J \rbrack  = N \int \lbrack {\cal D} C\rbrack
\lbrack {\cal D} \eta \rbrack
\lbrack {\cal D}\bar{\eta} \rbrack
exp \left\lbrace i \int {\rm d}^4 x ( {\cal L}
- \frac{1}{2 \alpha} \eta_{\alpha \beta} f^{\alpha} f^{\beta}
+ \bar{\eta} {\mathbf M } \eta
+ J^{\mu}_{\alpha} C^{\alpha}_{\mu}) \right\rbrace,
\label{10.34}
\ee
where $\int {\rm d}^4x \bar{\eta} {\mathbf M } \eta$ is a
simplified notation, whose explicit expression is
\be
\int {\rm d}^4x \bar{\eta} {\mathbf M } \eta
= \int {\rm d}^4 x {\rm d}^4 y ~
\bar{\eta}_{\alpha} (x) {\mathbf M}^{\alpha}_{~\beta}(x,y)
\eta^{\beta} (y).
\label{10.35}
\ee
The appearance of the non-trivial ghost fields is a inevitable
result of the non-Able nature of the gravitational gauge group.
\\

Set external source $J^{\mu}_{\alpha}$ to zero, we get,
\be
W\lbrack 0 \rbrack  = N \int \lbrack {\cal D} C\rbrack
\lbrack {\cal D} \eta \rbrack
\lbrack {\cal D}\bar{\eta} \rbrack
exp \left\lbrace i \int {\rm d}^4 x ( {\cal L}
- \frac{1}{2 \alpha} \eta_{\alpha \beta} f^{\alpha} f^{\beta}
+ \bar{\eta} {\mathbf M } \eta ) \right\rbrace,
\label{10.3501}
\ee
Now, let's take Lorentz covariant gauge condition,
\be
f^{\alpha} (C) = \partial^{\mu} C_{\mu}^{\alpha} .
\label{10.36}
\ee
Then
\be
\int {\rm d}^4x \bar{\eta} {\mathbf M } \eta =
- \int {\rm d}^4x \left( \partial^{\mu}
\bar{\eta}_{\alpha} (x) \right)
{\mathbf D}_{\mu~\beta}^{\alpha}(x) \eta^{\beta} (x).
\label{10.37}
\ee
And eq.(\ref{10.3501}) is changed into
\be
W\lbrack 0 \rbrack
 =  N \int \lbrack {\cal D} C\rbrack
\lbrack {\cal D} \eta \rbrack
\lbrack {\cal D}\bar{\eta} \rbrack
exp \left\lbrace i \int {\rm d}^4 x ( {\cal L}
- \frac{1}{2 \alpha}
\eta_{\alpha \beta} f^{\alpha} f^{\beta}
 - (\partial^{\mu}\bar{\eta}_{\alpha} )
{\mathbf D}_{\mu~\sigma}^{\alpha} \eta^{\sigma}
) \right\rbrace.
\label{10.3701}
\ee
For quantum gauge general relativity, the external source
of gravitational gauge field should be introduced in
a special way. Define the generating functional
with external sources as
\be
\begin{array}{rcl}
W\lbrack J, \beta, \bar{\beta} \rbrack
& = & N \int \lbrack {\cal D} C\rbrack
\lbrack {\cal D} \eta \rbrack
\lbrack {\cal D}\bar{\eta} \rbrack
exp \left\lbrace i \int {\rm d}^4 x ( {\cal L}
- \frac{1}{2 \alpha}
\eta_{\alpha \beta} f^{\alpha} f^{\beta} \right. \\
&&\\
&& \left. - (\partial^{\mu}\bar{\eta}_{\alpha} )
{\mathbf D}_{\mu~\sigma}^{\alpha} \eta^{\sigma}
+  C^{\alpha}_{\mu}
\swav{\delta}^{\mu\beta}_{\alpha\nu}(x)
J^{\nu}_{0\beta}
+ \bar{\eta}_{\alpha} \beta^{\alpha}
+ \bar{\beta}_{\alpha} \eta^{\alpha}
) \right\rbrace \\
&&\\
& = & N \int \lbrack {\cal D} C\rbrack
\lbrack {\cal D} \eta \rbrack
\lbrack {\cal D}\bar{\eta} \rbrack
exp \left\lbrace i \int {\rm d}^4 x ( {\cal L}
- \frac{1}{2 \alpha}
\eta_{\alpha \beta} f^{\alpha} f^{\beta} \right. \\
&&\\
&& \left. - (\partial^{\mu}\bar{\eta}_{\alpha} )
{\mathbf D}_{\mu~\sigma}^{\alpha} \eta^{\sigma}
+  C^{\alpha}_{\mu}
J^{\mu}_{\alpha}
+ \bar{\eta}_{\alpha} \beta^{\alpha}
+ \bar{\beta}_{\alpha} \eta^{\alpha}
) \right\rbrace ,
\end{array}
\label{10.38}
\ee
where $\swav{\delta}^{\mu \gamma}_{\alpha\rho} (x)$
is defined by
\be
\swav{\delta}^{\mu \gamma}_{\alpha\rho} (x)
\define \frac{1}{2}
\left (
\swav{\delta}^{\mu}_{\rho}(x)  \swav{\delta}^{\gamma}_{\alpha} (x)
+ \swav{\eta}^{\mu\gamma}(x)  \swav{\eta}_{\alpha\rho}(x)
\right),
\label{10.3801}
\ee
and
\be
J^{\mu}_{\alpha} \define
\swav{\delta}^{\mu\beta}_{\alpha\nu} (x)
J^{\nu}_{0\beta}.
\label{10.3802}
\ee
In the above definition, $\swav{\delta}^{\mu}_{\rho}(x)$,
$\swav{\eta}^{\mu\gamma}(x)$ and $\swav{\eta}_{\mu\gamma}(x)$
are defined by
\be
\swav{\delta}^{\mu}_{\rho}(x)
= \delta^{\mu}_{\rho}
- \frac{\partial^{\mu} \partial_{\rho}}
{\square  + i \epsilon},
\label{10.3803}
\ee
\be
\swav{\eta}^{\mu\gamma}(x)
= \eta^{\mu\gamma}
- \frac{\partial^{\mu} \partial^{\gamma}}
{\square  + i \epsilon},
\label{10.3804}
\ee
\be
\swav{\eta}_{\mu\gamma}(x)
= \eta_{\mu\gamma}
- \frac{\partial_{\mu} \partial_{\gamma}}
{\square  + i \epsilon},
\label{10.3805}
\ee
where
\be
\square \define \partial^2
= \partial^{\mu} \partial_{\mu}
= \eta^{\mu\nu} \partial_{\mu} \partial_{\nu}.
\label{10.3806}
\ee
Using these relations, we can prove that
\be
J^{\mu}_{\alpha} =
\swav{\delta}^{\mu\beta}_{\alpha\nu}
J^{\nu}_{\beta}.
\label{10.3807}
\ee
\\

The effective Lagrangian ${\cal L}_{eff}$
is defined by
\be
{\cal L}_{eff} \equiv
{\cal L} - \frac{1}{2 \alpha}
\eta_{\alpha \beta} f^{\alpha} f^{\beta}
- (\partial^{\mu}\bar{\eta}_{\alpha} )
{\mathbf D}_{\mu~\sigma}^{\alpha} \eta^{\sigma}.
\label{10.39}
\ee
${\cal L}_{eff}$ can be separate into free Lagrangian
${\cal L}_F$  and interaction Lagrangian ${\cal L}_I$,
\be
{\cal L}_{eff} = {\cal L}_F + {\cal L}_I,
\label{10.40}
\ee
where
\be
\begin{array}{rcl}
{\cal L}_F &=& - \frac{1}{16} \eta^{\mu \rho} \eta^{\nu \sigma} \eta_{\alpha \beta }
F_{0 \mu \nu}^{\alpha} F_{0 \rho \sigma}^{\beta}
- \frac{1}{8} \eta^{\mu \rho}
F^{\alpha}_{0 \mu \beta} F^{\beta}_{0 \rho \alpha}
+ \frac{1}{4} \eta^{\mu \rho}
F^{\alpha}_{0 \mu \alpha} F^{\beta}_{0 \rho \beta} \\
&&\\
&&  -\frac{1}{2 \alpha} \eta_{\alpha \beta}
(\partial^{\mu} C_{\mu}^{\alpha})
(\partial^{\nu} C_{\nu}^{\beta})
- (\partial^{\mu} \bar{\eta}_{\alpha})
(\partial_{\mu} \eta^{\alpha}),
\end{array}
\label{10.41}
\ee
\be
\begin{array}{rcl}
{\cal L}_I &=&  + g(\partial^{\mu} \bar{\eta}_{\alpha})
C_{\mu}^{\beta} (\partial_{\beta} \eta^{\alpha})
- g(\partial^{\mu} \bar{\eta}_{\alpha})
(\partial_{\sigma} C_{\mu}^{\alpha}) \eta^{\sigma}\\
&&\\
&& + {~ self ~interaction~ terms ~of ~Gravitational~gauge~field}.
\end{array}
\label{10.42}
\ee
From the interaction Lagrangian, we can see that ghost fields
do not couple to $J(C)$. This is the reflection of the fact
that ghost fields are not physical fields, they are virtual fields.
Besides, the gauge fixing term does not couple to $J(C)$ either.
Using effective Lagrangian ${\cal L}_{eff}$, the generating
functional $W\lbrack J, \beta, \bar{\beta} \rbrack$ can be
simplified to
\be
W\lbrack J, \beta, \bar{\beta} \rbrack
=  N \int \lbrack {\cal D} C\rbrack
\lbrack {\cal D} \eta \rbrack
\lbrack {\cal D}\bar{\eta} \rbrack
exp \left\lbrace i \int {\rm d}^4 x ( {\cal L}_{eff}
+ J^{\mu}_{\alpha} C^{\alpha}_{\mu}
+ \bar{\eta}_{\alpha} \beta^{\alpha}
+ \bar{\beta}_{\alpha} \eta^{\alpha}
) \right\rbrace,
\label{10.43}
\ee
\\

\section{Propagators}
\setcounter{equation}{0}

Using eq.(\ref{10.41}), we can deduce propagator of gravitational gauge
fields and ghost fields. First, after a partial integration,
 we change the form of
eq. (\ref{10.41}) into
\be
\int {\rm d}^4x {\cal L}_F =
\int {\rm d}^4x \left\lbrace \frac{1}{2}
C_{\mu}^{\alpha} {\mathbb M}^{\mu\nu}_{\alpha\beta} (x) C_{\nu}^{\beta}
+ \bar{\eta}_{\alpha}
\partial^2 \eta^{\alpha} \right\rbrace,
\label{10.44}
\ee
where the operator ${\mathbb M}^{\mu\nu}_{\alpha\beta} (x)$
is defined by
\be
\ba{rcl}
{\mathbb M}^{\mu\nu}_{\alpha\beta} (x) & = &
\frac{1}{4} \eta^{\mu\nu} \eta_{\alpha \beta}
\partial^{\rho} \partial_{\rho}
- \frac{1}{4} \eta_{\alpha \beta} (1 - \frac{4}{\alpha})
\partial^{\mu} \partial^{\nu}
- \frac{1}{4} \delta^{\mu}_{\beta}
\partial^{\nu} \partial_{\alpha} \\
&&\\
&& + \frac{1}{4} \delta^{\mu}_{\beta} \delta^{\nu}_{\alpha }
\partial^{\rho} \partial_{\rho}
- \frac{1}{4} \delta^{\nu}_{\alpha}
\partial^{\mu} \partial_{\beta}
+ \frac{1}{2} \delta^{\nu}_{\beta}
\partial^{\mu} \partial_{\alpha}  \\
&&\\
&& - \frac{1}{4} \eta^{\mu\nu}
\partial_{\alpha} \partial_{\beta}
- \frac{1}{2} \delta^{\mu}_{\alpha} \delta^{\nu}_{\beta}
\partial^{\rho} \partial_{\rho}
+ \frac{1}{2} \delta^{\mu}_{\alpha}
\partial^{\nu} \partial_{\beta}.
\ea
\label{10.4401}
\ee
\\

Denote the propagator of gravitational gauge field as
\be
-i \Delta_{F \mu \nu}^{\alpha \beta} (x),
\label{10.45}
\ee
and denote the propagator of ghost field as
\be
-i \Delta_{F \beta}^{\alpha } (x).
\label{10.46}
\ee
They satisfy the following equation,
\be
- {\mathbb M}^{\mu\nu}_{\alpha\beta} (x)
\Delta_{F \nu \rho}^{\beta \gamma} (x-y)
= \swav{\delta}^{\mu \gamma}_{\alpha\rho} (x)
\delta(x-y) ,
\label{10.47}
\ee
\be
- \partial^2 \Delta_{F \beta}^{\alpha } (x-y)
= \delta_{\beta}^{\alpha } \delta(x-y),
\label{10.48}
\ee
where $\swav{\delta}^{\mu \gamma}_{\alpha\rho} (x)$
is defined by (\ref{10.3801}).
\\

Make Fourier transformations to momentum space
\be
 - i \Delta_{F \mu \nu}^{\alpha \beta} (x)
= \int \frac{{\rm d}^4k}{(2 \pi)^4} ( - i )
\swav{\Delta}_{F \mu \nu}^{\alpha \beta}(k) \cdot  e^{ikx},
\label{10.49}
\ee
\be
- i \Delta_{F \beta}^{\alpha } (x)
= \int \frac{{\rm d}^4k}{(2 \pi)^4} ( - i )
\swav{\Delta}_{F \beta}^{\alpha }(k) \cdot e^{ikx},
\label{10.50}
\ee
where $- i \swav{\Delta}_{F \mu \nu}^{\alpha \beta}(k)$
and $ - i \swav{\Delta}_{F \beta}^{\alpha }(k)$ are corresponding
propagators in momentum space. They satisfy the following
equations,
\be
- {\mathbb M}^{\mu\nu}_{\alpha\beta} (k)
\swav{\Delta}_{F \nu \rho}^{\beta \gamma}(k)
= \swav{\delta}^{\mu\gamma}_{\alpha\rho} (k),
\label{10.51}
\ee
\be
k^2 \swav{\Delta}_{F \beta}^{\alpha }(k)
= \delta^{\alpha}_{\beta},
\label{10.52}
\ee
where the operator ${\mathbb M}^{\mu\nu}_{\alpha\beta} (k)$
is defined by
\be
\ba{rcl}
{\mathbb M}^{\mu\nu}_{\alpha\beta} (k)
& \define &
- \frac{1}{4} \eta^{\mu\nu} \eta_{\alpha \beta} k^2
+ \frac{1}{4} \eta_{\alpha \beta} (1 - \frac{4}{\alpha})
k^{\mu} k^{\nu}
+ \frac{1}{4} \delta^{\mu}_{\beta}
k^{\nu} k_{\alpha} \\
&&\\
&& - \frac{1}{4} \delta^{\mu}_{\beta} \delta^{\nu}_{\alpha }
k^2
+ \frac{1}{4} \delta^{\nu}_{\alpha}
k^{\mu} k_{\beta}
- \frac{1}{2} \delta^{\nu}_{\beta}
k^{\mu} k_{\alpha}  \\
&&\\
&& + \frac{1}{4} \eta^{\mu\nu}
k_{\alpha} k_{\beta}
+ \frac{1}{2} \delta^{\mu}_{\alpha} \delta^{\nu}_{\beta}
k^2
- \frac{1}{2} \delta^{\mu}_{\alpha}
k^{\nu} k_{\beta},
\ea
\label{10.5201}
\ee
and $\swav{\delta}^{\mu \gamma}_{\alpha\rho} (k)$ is defined by
\be
 \swav{\delta}^{\mu \gamma}_{\alpha\rho} (k) =
\frac{1}{2}
\left (
\swav{\delta}^{\mu}_{\rho} (k)  \swav{\delta}^{\gamma}_{\alpha} (k)
+ \swav{\eta}^{\mu\gamma} (k)  \swav{\eta}_{\alpha\rho} (k)
\right).
\label{10.5202}
\ee
The operator ${\mathbb M}^{\mu\nu}_{\alpha\beta}$
has the following symmetric property
\be
{\mathbb M}^{\mu\nu}_{\alpha\beta}
={\mathbb M}^{\nu\mu}_{\beta\alpha}.
\ee
In the above relation, $\swav{\delta}^{\mu}_{\rho}(k)$,
$\swav{\eta}^{\mu\gamma}(k)$ and $\swav{\eta}_{\mu\gamma} (k)$
are defined by
\be
\swav{\delta}^{\mu}_{\rho} (k)
= \delta^{\mu}_{\rho}
- \frac{k^{\mu} k_{\rho}}
{k^2  - i \epsilon},
\label{10.5203}
\ee
\be
\swav{\eta}^{\mu\gamma} (k)
= \eta^{\mu\gamma}
- \frac{k^{\mu} k^{\gamma}}
{k^2  - i \epsilon},
\label{10.5204}
\ee
\be
\swav{\eta}_{\mu\gamma} (k)
= \eta_{\mu\gamma}
- \frac{k_{\mu} k_{\gamma}}
{k^2  - i \epsilon}.
\label{10.5205}
\ee
It can be easily proved that $\swav{\delta}^{\mu}_{\rho} (k)$,
$\swav{\eta}^{\mu\gamma} (k)$, $\swav{\eta}_{\mu\gamma} (k)$,
${\delta}^{\mu}_{\rho} $,
${\eta}^{\mu\gamma} $ and ${\eta}_{\mu\gamma} $
satisfy the following relations:
\be
\swav{\eta}^{\mu\gamma} (k) \cdot
\swav{\eta}_{\gamma \nu} (k)
= {\eta}^{\mu\gamma} \cdot
\swav{\eta}_{\gamma \nu} (k)
= \swav{\eta}^{\mu\gamma} (k) \cdot
{\eta}_{\gamma \nu}
= \swav{\delta}^{\mu}_{\nu}(k),
\label{10.5208}
\ee
\be
\swav{\delta}^{\mu}_{\gamma} (k) \cdot
\swav{\delta}^{\gamma}_{ \nu} (k)
= {\delta}^{\mu}_{\gamma} \cdot
\swav{\delta}^{\gamma}_{ \nu} (k)
= \swav{\delta}^{\mu}_{\gamma} (k) \cdot
{\delta}^{\gamma}_{ \nu}
= \swav{\delta}^{\mu}_{\nu} (k),
\label{10.5211}
\ee
\be
\swav{\delta}^{\mu}_{\gamma} (k) \cdot
\swav{\eta}^{\gamma \nu} (k)
= {\delta}^{\mu}_{\gamma} \cdot
\swav{\eta}^{\gamma \nu} (k)
= \swav{\delta}^{\mu}_{\gamma} (k) \cdot
{\eta}^{\gamma \nu}
= \swav{\eta}^{\mu\nu} (k),
\label{10.5214}
\ee
\be
\swav{\delta}^{\gamma}_{\mu} (k) \cdot
\swav{\eta}_{\gamma \nu} (k)
= {\delta}^{\gamma}_{\mu} \cdot
\swav{\eta}_{\gamma \nu} (k)
= \swav{\delta}^{\gamma}_{\mu} (k) \cdot
{\eta}_{\gamma \nu}
= \swav{\eta}_{\mu\nu}(k),
\label{10.5217}
\ee
\be
k^{\mu} \swav{\eta}_{\mu \nu} (k)
= k_{\mu} \swav{\eta}^{\mu \nu} (k)
= k^{\mu} \swav{\delta}_{\mu}^{ \nu} (k)
= k_{\mu} \swav{\delta}^{\mu}_{ \nu} (k)
= 0.
\label{10.5221}
\ee
Using all these relations, we can prove that
$\swav{\delta}^{\mu \gamma}_{\alpha\rho} (k)$
satisfies the following relation
\be
\swav{\delta}^{\mu \gamma}_{\alpha\rho} (k)
\cdot \swav{\delta}^{\rho \beta}_{\gamma\nu} (k)
= \swav{\delta}^{\mu \beta}_{\alpha\nu} (k).
\label{10.5222}
\ee
\\

For the propagator of gravitational gauge field, we require that
it should satisfy the following gauge conditions
\be \label{10.53a}
\swav{\delta}_{\beta\rho}^{\nu\gamma} (x)
\cdot \Delta_{F \mu\nu}^{\alpha\beta} (x)
= \Delta_{F \mu\rho}^{\alpha\gamma} (x),
\ee
\be \label{10.53b}
\swav{\delta}_{\alpha\rho}^{\mu\gamma} (x)
\cdot \Delta_{F \mu\nu}^{\alpha\beta} (x)
= \Delta_{F \rho\nu}^{\gamma\beta} (x).
\ee
In momentum space, these two gauge conditions become
\be \label{10.53c}
 \swav\Delta_{F \mu\nu}^{\alpha\beta} (k)
 \cdot \swav{\delta}_{\beta\rho}^{\nu\gamma} (k)
= \swav\Delta_{F \mu\rho}^{\alpha\gamma} (k),
\ee
\be \label{10.53d}
\swav{\delta}_{\alpha\rho}^{\mu\gamma} (k)
\cdot \swav\Delta_{F \mu\nu}^{\alpha\beta} (k)
= \swav\Delta_{F \rho\nu}^{\gamma\beta} (k).
\ee
These two gauge conditions are related to the zero mass of graviton.
The solutions to the two propagator equations (\ref{10.51})
and (\ref{10.52}) and gauge conditions (\ref{10.53c} ) and
(\ref{10.53d})give out the propagators in momentum space,
\be
-i \swav{\Delta}_{F \mu \nu}^{\alpha \beta}(k)
= \frac{-i}{k^2 - i \epsilon}
\left\lbrack
\swav{\eta}_{\mu\nu}(k)  \swav{\eta}^{\alpha\beta}(k)
+ \swav{\delta}_{\mu}^{\beta}(k)  \swav{\delta}^{\alpha}_{\nu}(k)
- \swav{\delta}_{\mu}^{\alpha}(k)  \swav{\delta}_{\nu}^{\beta}(k)
\right\rbrack,
\label{10.53}
\ee
\be
-i \swav{\Delta}_{F \beta}^{\alpha }(k) =
\frac{-i}{k^2 - i \epsilon} \delta^{\alpha}_{\beta}.
\label{10.54}
\ee
The forms of these propagators are quite beautiful and symmetric.
it can be easily proved that
\be\label{10.54a}
k^{ \mu} \swav{\Delta}_{F \mu \nu}^{\alpha \beta}(k)
= k_{ \alpha} \swav{\Delta}_{F \mu \nu}^{\alpha \beta}(k)
= \swav{\Delta}_{F \mu \nu}^{\alpha \beta}(k) k^{ \nu}
= \swav{\Delta}_{F \mu \nu}^{\alpha \beta}(k) k_{ \beta}
= 0.
\ee
\\

\section{Feynman Rules of  Interaction Vertices}
\setcounter{equation}{0}

The interaction Lagrangian ${\cal L}_I$ is a function of
gravitational gauge field $C_{\mu}^{\alpha}$ and ghost
fields $\eta^{\alpha}$ and $\bar \eta_{\alpha}$,
\be
{\cal L}_I =
{\cal L}_I ( C, \eta, \bar\eta ).
\label{10.55}
\ee
Then eq.(\ref{10.43}) is changed into,
\be
\begin{array}{rcl}
W\lbrack J, \beta, \bar{\beta} \rbrack
& = & N \int \lbrack {\cal D} C\rbrack
\lbrack {\cal D} \eta \rbrack
\lbrack {\cal D}\bar{\eta} \rbrack
~exp \left\lbrace i \int {\rm d}^4 x
 {\cal L}_I ( C, \eta, \bar\eta ) \right\rbrace \\
&&\\
&&\cdot exp \left\lbrace i \int {\rm d}^4 x ( {\cal L}_F
+ J^{\mu}_{\alpha} C^{\alpha}_{\mu}
+ \bar{\eta}_{\alpha} \beta^{\alpha}
+ \bar{\beta}_{\alpha} \eta^{\alpha}
) \right\rbrace  \\
&&\\
&=& exp \left\lbrace i \int {\rm d}^4 x
 {\cal L}_I ( \frac{1}{i}\frac{\delta}{\delta J},
\frac{1}{i}\frac{\delta}{\delta \bar \beta},
\frac{1}{-i}\frac{\delta}{\delta \beta} ) \right\rbrace
\cdot W_0\lbrack J, \beta, \bar{\beta} \rbrack ,
\end{array}
\label{10.56}
\ee
where
\be
\begin{array}{rcl}
W_0\lbrack J, \beta, \bar{\beta} \rbrack
&=&  N \int \lbrack {\cal D} C\rbrack
\lbrack {\cal D} \eta \rbrack
\lbrack {\cal D}\bar{\eta} \rbrack
exp \left\lbrace i \int {\rm d}^4 x
\left( {\cal L}_F
+ J^{\mu}_{\alpha} C^{\alpha}_{\mu}
+ \bar{\eta}_{\alpha} \beta^{\alpha}
+ \bar{\beta}_{\alpha} \eta^{\alpha}
\right ) \right\rbrace  \\
&&\\
&=&  N \int \lbrack {\cal D} C\rbrack
\lbrack {\cal D} \eta \rbrack
\lbrack {\cal D}\bar{\eta} \rbrack
exp \left\lbrace i \int {\rm d}^4 x
\left( \frac{1}{2} C_{\mu}^{\alpha}
{\mathbb M}^{\mu\nu}_{\alpha\beta} (x) C_{\nu}^{\beta}
+ \bar{\eta}_{\alpha}
\partial^2 \eta^{\alpha}  \right. \right.  \\
&&\\
&& \left.\left.
+ J^{\mu}_{\alpha} C^{\alpha}_{\mu}
+ \bar{\eta}_{\alpha} \beta^{\alpha}
+ \bar{\beta}_{\alpha} \eta^{\alpha}
\right ) \right\rbrace  \\
&&\\
&=&  exp \left\lbrace
i \int\int {\rm d}^4 x {\rm d}^4 y
\left\lbrack  \frac{1}{2} J^{\mu}_{\alpha} (x)
\Delta_{F \mu \nu}^{\alpha \beta} (x-y)
J^{\nu}_{\beta} (y) \right. \right.  \\
&&\\
&& \left.\left.~~+ \bar\beta_{\alpha}(x)
\Delta_{F \beta}^{\alpha } (x-y) \beta^{\beta}(y)
\right\rbrack  \right\rbrace .
\end{array}
\label{10.57}
\ee
In order to obtain the above relation, eq. (\ref{10.3807})
is used.
\\

The interaction Feynman rules for interaction vertices
can be obtained from the interaction Lagrangian ${\cal L}_I$.
For example, the interaction
Lagrangian between gravitational gauge field and ghost field
is
\be
+ g(\partial^{\mu} \bar{\eta}_{\alpha})
C_{\mu}^{\beta} (\partial_{\beta} \eta^{\alpha})
- g(\partial^{\mu} \bar{\eta}_{\alpha})
(\partial_{\sigma} C_{\mu}^{\alpha}) \eta^{\sigma}.
\label{10.58}
\ee
This vertex belongs to
$C_{\mu}^{\alpha}(k) \bar{\eta}_{\beta}(-q) \eta^{\delta}(p) $
three body interactions, its Feynman rule is
\be
i g \delta^{\beta}_{\delta} q^{\mu} p_{\alpha}
- i g \delta^{\beta}_{\alpha} q^{\mu} k_{\delta}.
\label{10.59}
\ee
\\

To calculate the interaction lagrangian of three gravitational gauge
field, four  gravitational gauge field and higher  gravitational gauge
field are extremely complicated. Here I only explain how to calculate
then and list related results.
First, we can expand ${\rm det} G^{-1}$, $G^{-1 \nu}_{\alpha}$
and $g_{\alpha\beta}$ in terms of gravitational gauge field
\be\label{10.60}
{\rm det} G^{-1} = 1 + g C^{\alpha}_{\alpha}
+ \frac{g^2}{2} \left \lbrack  C_{\mu}^{\alpha} C_{\alpha}^{\mu}
+ C^{\mu}_{\mu} C_{\alpha}^{\alpha} \right \rbrack + \cdots ,
\ee
\be\label{10.61}
 G^{-1 \nu}_{\alpha} = \delta_{\alpha}^{\nu} + g C^{\nu}_{\alpha}
+ g^2    C^{\nu}_{\mu} C_{\alpha}^{\mu}
 + \cdots ,
\ee
\be\label{10.62}
\ba{rcl}
g_{\alpha\beta} & = & \eta_{\alpha\beta}
+ g \left \lbrack \eta_{\mu\beta} C^{\mu}_{\alpha}
+ \eta_{\mu\alpha} C^{\mu}_{\beta} \right \rbrack  \\
&&\\
&& + g^2 \left \lbrack \eta_{\mu\beta} C^{\mu}_{\alpha_1} C^{\alpha_1}_{\alpha}
+ \eta_{\mu\alpha} C^{\mu}_{\alpha_1} C^{\alpha_1}_{\beta}
+ \eta_{\mu\nu} C^{\mu}_{\alpha} C^{\nu}_{\beta} \right \rbrack
+ \cdots .
\ea
\ee
\\

Next, we need to expand the lagrangian ${\cal L}_0$ in terms of
gravitational gauge field. We will make the following expanding
\be\label{10.63}
{\cal L}_0 = \stackrel{2}{\cal L}_0 + \stackrel{3}{\cal L}_0
+ \stackrel{4}{\cal L}_0 + \cdots,
\ee
where $\stackrel{n}{\cal L}$ contains all n-th order interaction
terms of  gravitational gauge field. Substitute equations
(\ref{10.61}) and (\ref{10.62}) into (\ref{2.10}), we can get
\be\label{10.64}
\stackrel{2}{\cal L}_0 =
V_{\alpha\beta}^{\mu\nu\rho\sigma}
\left( \partial_{\rho} C_{\mu}^{\alpha} \right )
\left( \partial_{\sigma} C_{\nu}^{\beta} \right ),
\ee
where
\be\label{10.65}
V_{\alpha\beta}^{\mu\nu\rho\sigma} =
- \frac{1}{16} \bar{\eta}_{\alpha\beta}^{\mu\nu\rho\sigma}.
\ee
In the above relation, $\bar{\eta}_{\alpha\beta}^{\mu\nu\rho\sigma}$
is defined by
\be\label{10.66}
\bar{\eta}_{\alpha\beta}^{\mu\nu\rho\sigma}
= \eta^{\mu\nu} \bar{\eta}^{\rho\sigma}_{\alpha\beta}
+ \eta^{\rho\sigma} \bar{\eta}^{\mu\nu}_{\alpha\beta}
- \eta^{\mu\sigma} \bar{\eta}^{\rho\nu}_{\alpha\beta}
- \eta^{\rho\nu} \bar{\eta}^{\mu\sigma}_{\alpha\beta},
\ee
where
\be\label{10.67}
\bar{\eta}_{\alpha\beta}^{\mu\nu}
= \eta^{\mu\nu} \eta_{\alpha\beta}
+ 2 \delta^{\mu}_{\beta} \delta^{\nu}_{\alpha}
- 4 \delta^{\mu}_{\alpha} \delta^{\nu}_{\beta}.
\ee
The interaction term of three gravitational  gauge field in the
${\cal L}_0$ is
\be\label{10.68}
\stackrel{3}{\cal L}_0 =
V_{\alpha\beta\gamma}^{\mu\nu\lambda\rho\sigma}
C_{\lambda}^{\gamma}
\left( \partial_{\rho} C_{\mu}^{\alpha} \right )
\left( \partial_{\sigma} C_{\nu}^{\beta} \right ),
\ee
where
\be\label{10.69}
V^{\mu\nu\lambda\rho\sigma}_{\alpha\beta\gamma}
= \frac{g}{16}( \bar{\eta}_{\alpha\beta\gamma}^{\mu\nu\lambda\rho\sigma}
+ \bar{\eta}_{\beta\alpha\gamma}^{\nu\mu\lambda\sigma\rho} ),
\ee
\be\label{10.70}
\bar{\eta}_{\alpha\beta\gamma}^{\mu\nu\lambda\rho\sigma}
= \delta_{\gamma}^{\rho}
\bar{\eta}^{\mu\nu\lambda\sigma}_{\alpha\beta}
- \delta_{\beta}^{\lambda}
\bar{\eta}^{\mu\nu\rho\sigma}_{\alpha\gamma}.
\ee
The interaction term of four gravitational gauge field in the
${\cal L}_0$ is
\be\label{10.71}
\stackrel{4}{\cal L}_0 =
V_{\alpha\beta\gamma\delta}^{\mu\nu\lambda\kappa\rho\sigma}
C_{\lambda}^{\gamma} C_{\kappa}^{\delta}
\left( \partial_{\rho} C_{\mu}^{\alpha} \right )
\left( \partial_{\sigma} C_{\nu}^{\beta} \right ),
\ee
where
\be\label{10.72}
V_{\alpha\beta\gamma\delta}^{\mu\nu\lambda\kappa\rho\sigma}
= \frac{g^2}{64}
\left \lbrack
\bar{\eta}_{\alpha\beta\gamma\delta}^{\mu\nu\lambda\kappa\rho\sigma}
+ \bar{\eta}_{\beta\alpha\gamma\delta}^{\nu\mu\lambda\kappa\sigma\rho}
+ \bar{\eta}_{\alpha\beta\delta\gamma}^{\mu\nu\kappa\lambda\rho\sigma}
+ \bar{\eta}_{\beta\alpha\delta\gamma}^{\nu\mu\kappa\lambda\sigma\rho}
\right \rbrack.
\ee
\be\label{10.72a}
\bar{\eta}_{\alpha\beta\gamma\delta}^{\mu\nu\lambda\kappa\rho\sigma}
= \delta_{\beta}^{\kappa}
\bar{\eta}_{\alpha\delta\gamma}^{\mu\nu\lambda\rho\sigma}
+ \delta_{\delta}^{\rho}
\bar{\eta}_{\beta\gamma\alpha}^{\nu\mu\sigma\lambda\kappa}
+ \delta_{\beta}^{\lambda}
\bar{\eta}_{\alpha\gamma\delta}^{\mu\nu\kappa\rho\sigma}
+ \delta_{\alpha}^{\kappa}
\bar{\eta}_{\delta\beta\gamma}^{\mu\nu\lambda\rho\sigma}.
\ee
\\

Substitute above results and (\ref{10.60}) into (\ref{2.9}),
we get
\be\label{10.73}
\stackrel{3}{\cal L} =
\bar{V}_{\alpha\beta\gamma}^{\mu\nu\lambda\rho\sigma}
C_{\lambda}^{\gamma}
\left( \partial_{\rho} C_{\mu}^{\alpha} \right )
\left( \partial_{\sigma} C_{\nu}^{\beta} \right ),
\ee
where
\be\label{10.74}
\bar{V}^{\mu\nu\lambda\rho\sigma}_{\alpha\beta\gamma}
= {V}^{\mu\nu\lambda\rho\sigma}_{\alpha\beta\gamma}
- \frac{g}{16} \delta_{\gamma}^{\lambda}
\bar{\eta}_{\alpha\beta}^{\mu\nu\rho\sigma}.
\ee
And
\be\label{10.75}
\stackrel{4}{\cal L} =
\bar{V}_{\alpha\beta\gamma\delta}^{\mu\nu\lambda\kappa\rho\sigma}
C_{\lambda}^{\gamma} C_{\kappa}^{\delta}
\left( \partial_{\rho} C_{\mu}^{\alpha} \right )
\left( \partial_{\sigma} C_{\nu}^{\beta} \right ),
\ee
where
\be\label{10.76}
\ba{rcl}
\bar{V}_{\alpha\beta\gamma\delta}^{\mu\nu\lambda\kappa\rho\sigma}
&=& {V}_{\alpha\beta\gamma\delta}^{\mu\nu\lambda\kappa\rho\sigma}
- \frac{g^2}{32}
\left ( \delta_{\gamma}^{\kappa} \delta_{\delta}^{\lambda}
+ \delta_{\gamma}^{\lambda} \delta_{\delta}^{\kappa}
 \right ) \bar{\eta}_{\alpha\beta}^{\mu\nu\rho\sigma} \\
 &&\\
&&
+ \frac{g^2}{32}
\left (
\delta_{\delta}^{\kappa} \delta_{\gamma}^{\rho}
\bar{\eta}_{\alpha\beta}^{\mu\nu\lambda\sigma}
+ \delta_{\delta}^{\kappa} \delta_{\gamma}^{\sigma}
\bar{\eta}_{\beta\alpha}^{\nu\mu\lambda\rho}
+ \delta_{\gamma}^{\lambda} \delta_{\delta}^{\rho}
\bar{\eta}_{\alpha\beta}^{\mu\nu\kappa\sigma}
+ \delta_{\gamma}^{\lambda} \delta_{\delta}^{\sigma}
\bar{\eta}_{\beta\alpha}^{\nu\mu\kappa\rho}
 \right ) \\
 &&\\
&&
- \frac{g^2}{32}
\left (
 \delta_{\delta}^{\kappa} \delta_{\beta}^{\lambda}
\bar{\eta}_{\alpha\gamma}^{\mu\nu\rho\sigma}
+ \delta_{\delta}^{\kappa} \delta_{\alpha}^{\lambda}
\bar{\eta}_{\beta\gamma}^{\nu\mu\sigma\rho}
+ \delta_{\gamma}^{\lambda} \delta_{\beta}^{\kappa}
\bar{\eta}_{\alpha\delta}^{\mu\nu\rho\sigma}
+ \delta_{\gamma}^{\lambda} \delta_{\alpha}^{\kappa}
\bar{\eta}_{\beta\delta}^{\nu\mu\sigma\rho}
\right )
 .
 \ea
\ee
\\

Feynman rules for the vertex of three gravitational gauge field
$C_{\mu}^{\alpha} (p_1) C_{\nu}^{\beta} (p_2) C_{\lambda}^{\gamma} (p_3)$
is
\be\label{10.77}
- 2 i  \left\lbrack
\bar{V}^{\mu\nu\lambda\rho\sigma}_{\alpha\beta\gamma} p_{1 \rho} p_{2 \sigma}
+ \bar{V}^{\nu\lambda\mu\rho\sigma}_{\beta\gamma\alpha} p_{2 \rho} p_{3 \sigma}
+ \bar{V}^{\lambda\mu\nu\rho\sigma}_{\gamma\alpha\beta} p_{3 \rho} p_{1 \sigma}
\right\rbrack .
\ee
The Feynman rule for the vertex of four gravitational gauge field
$C_{\mu}^{\alpha} (p_1) C_{\nu}^{\beta} (p_2)
C_{\lambda}^{\gamma} (p_3) C_{\kappa}^{\delta} (p_4)$
is
\be\label{10.78}
\ba{rl}
-  4 i & \left\lbrack
\bar{V}^{\mu\nu\lambda\kappa\rho\sigma}_{\alpha\beta\gamma\delta} p_{1 \rho} p_{2 \sigma}
+ \bar{V}^{\mu\lambda\nu\kappa\rho\sigma}_{\alpha\gamma\beta\delta} p_{1 \rho} p_{3 \sigma}
+ \bar{V}^{\mu\kappa\nu\lambda\rho\sigma}_{\alpha\delta\beta\gamma} p_{1 \rho} p_{4 \sigma}
 \right .
\\
\\ & \left .
+ \bar{V}^{\nu\lambda\kappa\mu\rho\sigma}_{\beta\gamma\delta\alpha} p_{2 \rho} p_{3 \sigma}
+ \bar{V}^{\nu\kappa\mu\lambda\rho\sigma}_{\beta\delta\alpha\gamma} p_{2 \rho} p_{4 \sigma}
+ \bar{V}^{\lambda\kappa\mu\nu\rho\sigma}_{\gamma\delta\alpha\beta} p_{3 \rho} p_{4 \sigma}
\right\rbrack .
\ea
\ee
\\

\section{Discussions}
\setcounter{equation}{0}

In this paper, path integral quantization of quantum gauge general relativity
is discussed, and Feynman rules of various interaction vertices are calculated.
These results are needed in the loop diagram calculation. \\

In the literature \cite{06}, we have formally proved that
quantum gauge general relativity is a perturbatively renormalizable quantum
theory. In that proof, detailed calculations of loop diagrams are not
performed. In the next step, we will calculate all divergent one-loop diagrams,
discuss renormalization of quantum gauge general relativity in one-loop level,
and determine the renormalization constant in one-loop level. These results
will be summarize in the further paper. \\

{\bf [Acknowledgement]} The author would like to thank Prof. J.P. Hsu for
useful discussions and kindly suggestions on this work.
\\

\end{document}